\documentclass{aa}
\usepackage[varg]{txfonts}
\usepackage{graphicx,times}
\usepackage{natbib}
\usepackage{amssymb,amsmath}
\usepackage{txfonts,booktabs,threeparttable}
\usepackage[dvipsnames,usenames]{color}
\bibpunct{(}{)}{;}{a}{}{,}

\begin{document}

\title{The environmental dependence of the stellar mass fundamental plane of early-type galaxies}

   \author{Lei Hou\inst{1} \and Yu Wang\inst{1}
   }
+

   \institute{Department of Astronomy, University of Science and Technology of China, Jinzhai Road 96, Hefei 230026, China; {\it wywa@ustc.edu.cn}\\
   }

   \date{Received~~xxxxxx /
    accepted~~xxxxxx}

\abstract{}
{We investigate the environmental dependence of the stellar mass fundamental plane (FP$_*$) using the early-type galaxy sample from the Sloan Digital Sky Survey Data Release 7 (SDSS DR7).}
{The FP$_*$ is calculated by replacing the luminosity in the fundamental plane (FP) with stellar mass. Based on the SDSS group catalog, we characterize the galaxy environment according to the mass of the host dark matter halo and the position in the halo. In halos with the same mass bin, the color distributions of central and satellite galaxies are different. Therefore, we calculate FP$_*$ coefficients of galaxies in different environments and compare them with those of the FP to study the contribution of the stellar population.}
{We find that coefficient $a$ of the FP$_*$ is systematically larger than that of the FP, but coefficient $b$ of the FP$_*$ is similar to the FP. Moreover, the environmental dependence of the FP$_*$ is similar to that of the FP. For central galaxies, FP$_*$ coefficients are significantly dependent on the halo mass. For satellite galaxies, the correlation between FP$_*$ coefficients and the halo mass is weak.}
{We conclude that the tilt of the FP is not primarily driven by the stellar population.}

\titlerunning{Environmental Dependence of the Stellar Mass Fundamental Plane}
\authorrunning{L. Hou and Y. Wang}
\maketitle

\section{Introduction}
The fundamental plane (FP) is a scaling relation of early-type galaxies (ETGs) between the central velocity dispersion $\sigma_0$, the effective radius $R_0$, and the average surface brightness in the effective radius $I_0$ \citep{1987ApJ...313...59D,1987ApJ...313...42D}. It is written as 
\begin{equation}
\log R_0=a\log\sigma_0+b\log I_0+c.
\end{equation}
The scatter of the FP is quite small. Based on the assumption of structural and dynamical homology, ETGs have similar density and orbital distributions. If ETGs also have similar dark matter fraction and the stellar mass-to-light ratio, and assuming that the virial theorem holds at all radii, a FP with $a=2$ and $b=-1$ is expected. However, observed FP coefficients are found to differ from the virial prediction. This is called the tilt of the FP. Many studies have discussed the origin of the FP tilt, such as the structural and dynamical non-homology \citep{1997MNRAS.287..221G, 1997A&A...320..415B, 2004ApJ...600L..39T}, the variations of the stellar population \citep{1996A&A...309..749P, 1998ApJ...508L..43F}, and the fraction of the dark matter to the total mass \citep{1996MNRAS.282....1C, 2003MNRAS.341.1109B, 2004NewA....9..329P}. The conclusions in these studies often conflict with each other and there are also suggestions that the tilt of the FP is due to the contribution of multiple factors \citep{2013MNRAS.435...45D}.

Many works indicate that there is a systematic variation of stellar population parameters through the FP, such as $M_*/L$, age, and metallicity \citep{2009ASPC..419...96G,2012MNRAS.420.2773S}. Therefore, the stellar population seems to make an important contribution to the FP tilt. To resolve the contribution of the stellar population, \citet{2009MNRAS.396.1171H} measure the stellar mass fundamental plane (hereafter FP$_*$) by replacing the luminosity in the FP with the stellar mass
\begin{equation}
\log R_0=a\log\sigma_0+b\log \Sigma_0+c,
\end{equation}
where $\Sigma_0=M_*/(2\pi R_0^2)$ is the stellar mass surface density. The difference between FP and FP$_*$ is due to the stellar population, and the difference between FP$_*$ and the virial prediction is due to the non-homology and dark matter fraction. As a result, the contribution of stellar population to the FP tilt is resolved. \citet{2009MNRAS.396.1171H} compared the FP and FP$_*$ of ETGs in the Sloan Digital Sky Survey (SDSS), and found that the FP$_*$ is steeper than the FP, but shallower than the virial prediction. This means that the stellar population only contributes to a part of the FP tilt.

The properties of galaxies have been found to correlate with the environment. For example, the propotion of ETGs is larger in denser environment, which is called the morphology-density relation \citep{1980ApJ...236..351D}. Several studies have focused on the environmental dependence of the FP and found that FP coefficients in denser environments are different to those in lower density environments \citep{1991MNRAS.249..755L, 1992ApJ...389L..49D, 2003AJ....125.1866B, 2006AJ....131.1288B, 2008ApJ...685..875D, 2010MNRAS.408.1361L, 2012MNRAS.427..245M, 2013MNRAS.432.1709C}. However, the definitions of the environment in these studies are different and this makes it difficult to compare the conclusions. The SDSS group catalog \citep{2007ApJ...671..153Y} provides an approach for studying the correlation between galaxies and the environment. Each galaxy in the catalog is assigned to a dark matter halo and identified as a central or satellite galaxy in the halo. Based on the group catalog, \citet{2015RAA....15..651H} investigated the environmental dependence of the FP of ETGs in the SDSS, and found that FP coefficients of central galaxies are significantly correlated to the halo mass, but those of satellite galaxies are independent of the halo mass. The discrepancy between central and satellite galaxies is significant in small halos. However, central and satellite galaxies with the same halo mass have different color distribution. The environmental dependence of FP coefficients may be due to the variation of the color. We therefore need to test whether the environmental dependence of FP coefficients is due to the variation of the color. The FP$_*$ is the most useful tool to study the effect of the color.

The aim of this paper is to study the environmental dependence of the FP$_*$. To do this, we select the ETG sample from the Sloan Digital Sky Survey Data Release 7 (SDSS DR7), and use the SDSS group catalog to find the halo mass and the position in the halo for each galaxy. Then we measure FP$_*$ coefficients of ETGs in different environments. By comparing the environmental dependence of the FP and the FP$_*$, we can resolve the effect of the stellar population.

The paper is organized as follows. Section 2 describes the formulation of the FP and the FP$_*$. Section 3 introduces the ETG sample, the SDSS group catalog and the calculation of physical parameters. Section 4 describes the fitting method of the FP$_*$. We investigate the environmental dependence of the FP$_*$ in Sect.~5. We summarize this paper and discuss the results in Sect.~6. Throughout this paper, we adopt a $\Lambda$CDM cosmology with parameter $\Omega_m=0.238$, $\Omega_\Lambda=0.762$ and $h=0.73$.

\section{From the FP to the FP$_*$}

If the virial theorem holds at all radii, ETGs should satisfy
\begin{equation}
V^2\propto\frac{GM_{dyn}}{R}\propto \frac{M_{dyn}}{L}RI,
\end{equation}
where $M_{dyn}$ is the dynamical mass, $V\equiv\sqrt{v_c^2/2+\sigma^2}$, $v_c$ is the rotational velocity and $\sigma$ is the velocity dispersion. Assuming $V/\sigma$ is constant for all ETGs, then
\begin{equation}
\sigma^2\propto\frac{GM_{dyn}}{R}\propto \frac{M_{dyn}}{L}RI.
\label{equa:virial}
\end{equation}
If $M_{dyn}/L$ is also constant, the FP with $a=2$ and $b=-1$ is expected. The tilt of the FP is the evidence on the systematic variation of $M_{dyn}/L$. The dynamical mass-to-light ratio $M_{dyn}/L$ can be resolved into three factors
\begin{equation}
\frac{M_{dyn}}{L}=\frac{M_{dyn}}{M_{tot}}\frac{M_{tot}}{M_*}\frac{M_*}{L},
\end{equation}
where $M_{tot}$ is the total mass of dark matter and baryonic matter \citep{2009MNRAS.396.1171H}. Each factor corresponds to an explanation to the FP tilt: $M_{dyn}/M_{tot}$ represents the non-homology, $M_{tot}/M_*$ represents the fraction of the stellar mass to the total mass, and $M_*/L$ represents the stellar population. 

To examine the contribution of the stellar population to the FP, \citet{2009MNRAS.396.1171H} replaced $L$ into $M_*$ to produce the FP$_*$. Similar to Equation \ref{equa:virial}, the virial theorem is expressed as
\begin{equation}
\sigma^2\propto\frac{GM_{dyn}}{R}\propto \frac{M_{dyn}}{M_*}R\Sigma.
\label{equa:star}
\end{equation}
If $M_{dyn}/M_*$ is constant, the FP$_*$ satisfies the virial prediction. The effect of stellar population is eliminated from the FP$_*$.

\begin{figure*}
\begin{center}
\includegraphics[scale=0.9]{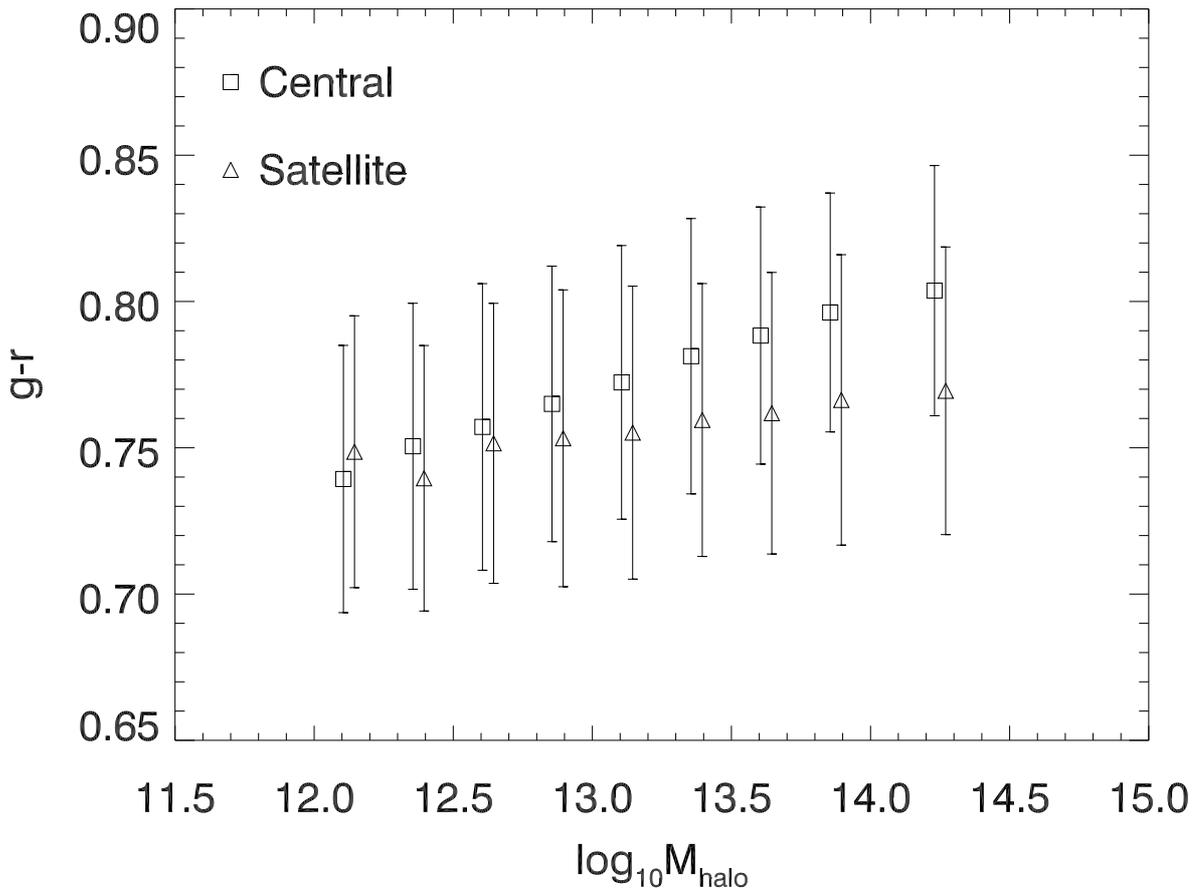}
\end{center}
\caption{The dependence of $g-r$ color on the halo mass for central and satellite galaxies. Squares are central galaxies, and triangles are satellite galaxies. Error bars indicate the standard deviation.}
\label{fig:bin_color}
\end{figure*}

In our work, we calculate the dependence of FP coefficients on the halo mass of central and satellite galaxies. However, in the same halo mass bin, the color distributions of central and satellite galaxies are different. In Fig.~\ref{fig:bin_color}, we plot $g-r$ color as a function of the halo mass for central and satellite galaxies respectively. We find that central galaxies in more massive halos are redder, but the color of satellite galaxies is only weakly correlated to the halo mass. This means that the environmental dependence of FP coefficients may be caused by the difference of the color distribution between central and satellite galaxies. Therefore, the environmental dependence of the FP$_*$ needs to be calculated, and compared with that of the FP to find the effect of the color.

\section{Data}
\subsection{Early-type galaxy sample}

ETGs in this work are selected from SDSS DR7\footnote{http://www.sdss.org/dr7/} using similar criteria to \citet{2003AJ....125.1817B}: 

(1) Concentration parameter $r_{90}/r_{50}>2.5$ in $i$ band; 

(2) Ratio of the likelihood of the de Vaucouleurs model to the exponential model $L_{deV}/L_{exp}\geq 1.03$; 

(3) Spectral classification index $eClass<-0.1$; 

(4) Warning flag $zWarning=0$; 

(5) $S/N>10$; 

(6) Redshift $z<0.2$; 

(7) Central velocity dispersion $\sigma>70~km~s^{-1}$. 

The sample consists of 70 793 galaxies. 

\subsection{SDSS group catalog}
The selection procedure and the properties of the group catalog are described in \citet{2007ApJ...671..153Y}. A modified version of the halo-based group finder that was developed by \citet{2005MNRAS.356.1293Y} is used to select galaxy groups from the SDSS. In this work, the catalog version based on the SDSS DR7 is used. By cross-correlating the ETG sample with the group catalog, we get the mass of the host dark matter halo for each ETG. We also distinguish each ETG as a central or satellite galaxy of the halo. In this work, we adopt the halo mass and whether a central or satellite galaxy to describe the environment of an ETG. The unit of the halo mass is $h^{-1}M_{\odot}$.

\subsection{Parameters}
We compute the effective angular radius as 
\begin{equation}
r_0=\sqrt{b/a}~r_{deV},
\end{equation}
where $b/a$ is the axis ratio and $r_{deV}$ is the de Vaucouleurs angular radius. Next, we convert $r_0$ into the effective physical radius $R_0$ at the corresponding redshift $z$. 

Because the SDSS spectrum is observed using a fixed fiber aperture, the velocity dispersion $\sigma$ should be corrected as 
\begin{equation}
\sigma_0=\sigma(\frac{r_{fiber}}{r_0/8})^{0.04},
\end{equation}
where $r_{fiber}$ is the angular radius of the fiber, and $r_0$ is the effective angular radius \citep{1995MNRAS.276.1341J,1999MNRAS.305..259W}.

To fit the FP$_*$ of ETGs, we should calculate the stellar mass $M_*$ of each galaxy. In the SDSS group catalog, $M_*$ is estimated based on the relation between the stellar mass-to-light ratio and the $g-r$ color \citep{2003ApJS..149..289B}:  
\begin{equation}
\log M_*/L_r=1.097(g-r)-0.306,
\end{equation}
where the magnitudes are K-corrected and evolution-corrected to $z=0.0$. Therefore, $M_*$ is estimated as
\begin{equation}
\log \frac{M_*}{h^{-2}M_\odot}=-0.306+1.097(g-r)-0.1-0.4(M_r-5\log h-4.64),
\end{equation}
where 4.64 is the r-band AB magnitude of the Sun, and -0.1 means that the \citet{2001MNRAS.322..231K} initial mass function (IMF) is adopted \citep{2006A&A...453..869B}. The stellar mass surface density is computed as $\Sigma_0=M_*/(2\pi R_0^2)$. In Section 6.2, we will discuss the effect of the IMF by changing the Kroupa IMF to another universal IMF, and to the non-universal IMF.

\section{Fitting methods}
In this work, we use the orthogonal fitting method to calculate coefficients of the FP$_*$. This fitting method is similar to that in \citet{2015RAA....15..651H}, except that $I_0$ is replaced with $\Sigma_0$. Firstly, we calculate the covariance matrix of $\log R_0$, $\log \sigma_0$, and $\log \Sigma_0$. Then we estimate the corresponding error matrix, and subtract it from the covariance matrix to get the intrinsic covariance matrix. When calculating the error matrix, we assume that the error of stellar mass is not correlated to the error of effective radius and velocity dispersion. By diagonalizing the intrinsic covariance matrix, we obtain the slopes of FP$_*$. The offset of FP$_*$ is calculated by $c=\overline{\log R_0}-a\overline{\log \sigma_0}-b\overline{\log \Sigma_0}$. We use 100 repeated bootstrap samples to estimate the uncertainties of $a$, $b$, and $c$.

With this method, we measure FP$_*$ coefficients of the ETG sample. Results are shown in Table \ref{tab:FPstar}. We find that, compared to the FP (Table 1 of \citet{2015RAA....15..651H}), coefficient $a$ of FP$_*$ is closer to 2, but coefficient $b$ of FP$_*$ is similar to that of the FP. This implies that the stellar population can only explain a part of the FP tilt. Therefore, $M_{dyn}/M_*$ is not constant.

\begin{table*}
\scriptsize
\begin{center}
\caption[]{FP$_*$ coefficients of the ETG sample \label{tab:FPstar}}
 \begin{tabular}{cccccc}
  \hline\noalign{\smallskip}
Band & $a$ & $b$ & $c$ & $scatter_{orth}$ & $scatter_{R_0}$\\
  \hline\noalign{\smallskip}
$g$ & $1.511\pm 0.004$ & $-0.716\pm 0.001$ & $3.32\pm 0.01$ & 0.052 & 0.102\\
$r$ & $1.503\pm 0.004$ & $-0.751\pm 0.001$ & $3.65\pm 0.02$ & 0.051 & 0.099\\
$i$ & $1.508\pm 0.004$ & $-0.766\pm 0.001$ & $3.78\pm 0.02$ & 0.050 & 0.098\\
$z$ & $1.479\pm 0.004$ & $-0.799\pm 0.002$ & $4.14\pm 0.02$ & 0.049 & 0.095\\
  \noalign{\smallskip}\hline
\end{tabular}
\end{center}
\tablefoot{
FP$_*$ coefficients of the ETG sample in $g$, $r$, $i$, and $z$ bands. Columns 2 to 4 are coefficients $a$, $b$, and $c$. Columns 5 and 6 are scatters in the orthogonal direction and in the $R_0$ direction.
}
\end{table*}

\section{Environmental dependence of the FP$_*$}
\begin{figure*}
\begin{center}
\includegraphics[scale=0.55]{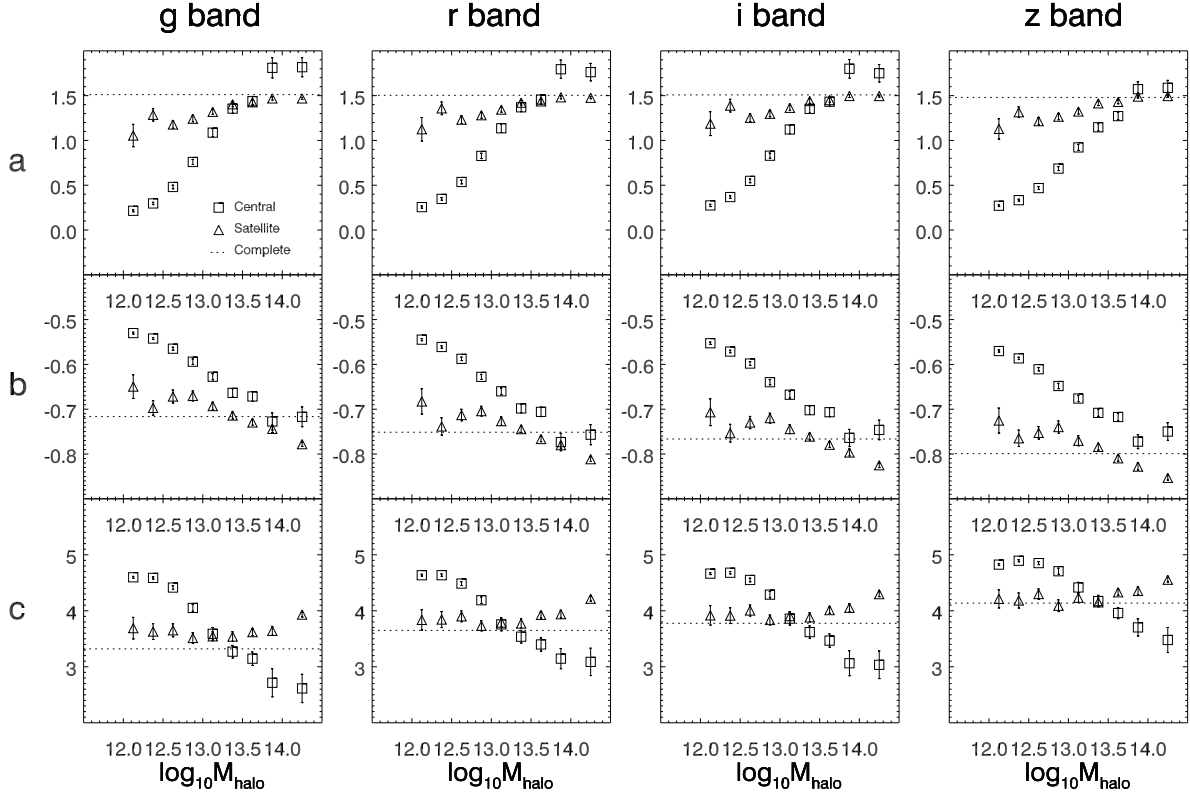}
\end{center}
\caption{Environmental dependence of the FP$_*$ in $g$, $r$, $i$ and $z$ bands. Squares are central galaxies, and triangles are satellite galaxies. The horizontal dotted lines are the corresponding coefficients of the complete ETG sample.}
\label{fig:fp_bin_star}
\end{figure*}

\citet{2015RAA....15..651H} studied the environmental dependence of the FP, and found that there is a significant difference between central and satellite galaxies. FP coefficients of central galaxies depend strongly on the halo mass but, for satellite galaxies, the dependence on the halo mass is weak. In this work, we study the environmental dependence on the FP$_*$. To do this, we divide the ETG sample into nine halo mass bins (as per Table 2 of \citet{2015RAA....15..651H}), and fit the FP$_*$ for each subsample. The FP$_*$ coefficients, as functions of the halo mass, are shown in Fig.~\ref{fig:fp_bin_star}. We compare the environmental dependence of the FP$_*$ and the FP in Tables \ref{tab:envcen} and \ref{tab:envsat} for central and satellite galaxies respectively. 

\begin{table*}[p]
\begin{center}
\caption[]{Environmental dependence of the FP and the FP$_*$ of central galaxies\label{tab:envcen}}
 \begin{tabular}{cccccccc}
  \hline\noalign{\smallskip}
Band & $log_{10}M_{halo}$ & \multicolumn{3}{c}{FP$_*$ Coefficients} & \multicolumn{3}{c}{FP Coefficients}\\
 & & $a$ & $b$ & $c$ & $a$ & $b$ & $c$\\
  \hline\noalign{\smallskip}
$g$ & $(12.00, 12.25]$ & $0.215\pm 0.010$ & $-0.530\pm 0.002$ & $4.60\pm 0.02$ & $0.075\pm 0.014$ & $-0.574\pm 0.003$ & $-4.28\pm 0.05$\\
  & $(12.25, 12.50]$ & $0.297\pm 0.015$ & $-0.542\pm 0.002$ & $4.59\pm 0.03$ & $0.109\pm 0.020$ & $-0.586\pm 0.004$ & $-4.38\pm 0.07$\\
  & $(12.50, 12.75]$ & $0.481\pm 0.021$ & $-0.565\pm 0.003$ & $4.41\pm 0.04$ & $0.265\pm 0.027$ & $-0.600\pm 0.004$ & $-4.80\pm 0.09$\\
  & $(12.75, 13.00]$ & $0.760\pm 0.026$ & $-0.594\pm 0.006$ & $4.05\pm 0.05$ & $0.642\pm 0.046$ & $-0.639\pm 0.006$ & $-5.96\pm 0.14$\\
  & $(13.00, 13.25]$ & $1.087\pm 0.045$ & $-0.628\pm 0.006$ & $3.59\pm 0.11$ & $1.046\pm 0.061$ & $-0.669\pm 0.008$ & $-7.14\pm 0.19$\\
  & $(13.25, 13.50]$ & $1.355\pm 0.048$ & $-0.664\pm 0.009$ & $3.27\pm 0.10$ & $1.203\pm 0.062$ & $-0.685\pm 0.008$ & $-7.64\pm 0.19$\\
  & $(13.50, 13.75]$ & $1.438\pm 0.051$ & $-0.672\pm 0.011$ & $3.14\pm 0.13$ & $1.215\pm 0.065$ & $-0.692\pm 0.010$ & $-7.73\pm 0.21$\\
  & $(13.75, 14.00]$ & $1.811\pm 0.128$ & $-0.727\pm 0.020$ & $2.71\pm 0.30$ & $1.475\pm 0.083$ & $-0.738\pm 0.017$ & $-8.73\pm 0.31$\\
  & $(14.00, \infty)$ & $1.817\pm 0.108$ & $-0.717\pm 0.022$ & $2.61\pm 0.29$ & $1.389\pm 0.078$ & $-0.731\pm 0.016$ & $-8.47\pm 0.26$\\
  \noalign{\smallskip}\hline
$r$ & $(12.00, 12.25]$ & $0.255\pm 0.011$ & $-0.545\pm 0.002$ & $4.63\pm 0.02$ & $0.132\pm 0.013$ & $-0.569\pm 0.003$ & $-4.19\pm 0.05$\\
  & $(12.25, 12.50]$ & $0.348\pm 0.017$ & $-0.561\pm 0.003$ & $4.64\pm 0.03$ & $0.187\pm 0.015$ & $-0.585\pm 0.003$ & $-4.37\pm 0.06$\\
  & $(12.50, 12.75]$ & $0.536\pm 0.022$ & $-0.588\pm 0.004$ & $4.49\pm 0.04$ & $0.340\pm 0.022$ & $-0.601\pm 0.004$ & $-4.79\pm 0.08$\\
  & $(12.75, 13.00]$ & $0.827\pm 0.027$ & $-0.628\pm 0.006$ & $4.19\pm 0.05$ & $0.673\pm 0.037$ & $-0.641\pm 0.005$ & $-5.84\pm 0.12$\\
  & $(13.00, 13.25]$ & $1.136\pm 0.048$ & $-0.660\pm 0.007$ & $3.76\pm 0.11$ & $1.013\pm 0.046$ & $-0.671\pm 0.007$ & $-6.86\pm 0.15$\\
  & $(13.25, 13.50]$ & $1.370\pm 0.048$ & $-0.698\pm 0.009$ & $3.54\pm 0.10$ & $1.185\pm 0.053$ & $-0.689\pm 0.008$ & $-7.40\pm 0.16$\\
  & $(13.50, 13.75]$ & $1.456\pm 0.051$ & $-0.706\pm 0.011$ & $3.40\pm 0.12$ & $1.247\pm 0.062$ & $-0.699\pm 0.010$ & $-7.63\pm 0.19$\\
  & $(13.75, 14.00]$ & $1.795\pm 0.121$ & $-0.773\pm 0.020$ & $3.15\pm 0.24$ & $1.499\pm 0.079$ & $-0.752\pm 0.017$ & $-8.66\pm 0.30$\\
  & $(14.00, \infty)$ & $1.762\pm 0.100$ & $-0.757\pm 0.023$ & $3.09\pm 0.27$ & $1.420\pm 0.074$ & $-0.744\pm 0.016$ & $-8.40\pm 0.25$\\
  \noalign{\smallskip}\hline
$i$ & $(12.00, 12.25]$ & $0.275\pm 0.011$ & $-0.553\pm 0.002$ & $4.66\pm 0.02$ & $0.198\pm 0.013$ & $-0.576\pm 0.003$ & $-4.31\pm 0.05$\\
  & $(12.25, 12.50]$ & $0.368\pm 0.017$ & $-0.571\pm 0.003$ & $4.68\pm 0.03$ & $0.269\pm 0.014$ & $-0.595\pm 0.003$ & $-4.55\pm 0.05$\\
  & $(12.50, 12.75]$ & $0.549\pm 0.022$ & $-0.598\pm 0.004$ & $4.55\pm 0.04$ & $0.425\pm 0.021$ & $-0.612\pm 0.004$ & $-4.98\pm 0.07$\\
  & $(12.75, 13.00]$ & $0.830\pm 0.026$ & $-0.640\pm 0.006$ & $4.29\pm 0.05$ & $0.732\pm 0.033$ & $-0.654\pm 0.005$ & $-5.99\pm 0.11$\\
  & $(13.00, 13.25]$ & $1.122\pm 0.048$ & $-0.668\pm 0.007$ & $3.86\pm 0.11$ & $1.023\pm 0.040$ & $-0.681\pm 0.007$ & $-6.86\pm 0.13$\\
  & $(13.25, 13.50]$ & $1.352\pm 0.047$ & $-0.702\pm 0.009$ & $3.62\pm 0.10$ & $1.190\pm 0.048$ & $-0.701\pm 0.007$ & $-7.40\pm 0.15$\\
  & $(13.50, 13.75]$ & $1.433\pm 0.050$ & $-0.707\pm 0.011$ & $3.47\pm 0.12$ & $1.266\pm 0.058$ & $-0.704\pm 0.010$ & $-7.60\pm 0.18$\\
  & $(13.75, 14.00]$ & $1.800\pm 0.121$ & $-0.764\pm 0.020$ & $3.06\pm 0.26$ & $1.508\pm 0.076$ & $-0.761\pm 0.016$ & $-8.64\pm 0.28$\\
  & $(14.00, \infty)$ & $1.750\pm 0.101$ & $-0.746\pm 0.020$ & $3.04\pm 0.28$ & $1.437\pm 0.071$ & $-0.747\pm 0.015$ & $-8.36\pm 0.24$\\
  \noalign{\smallskip}\hline  
$z$ & $(12.00, 12.25]$ & $0.271\pm 0.007$ & $-0.570\pm 0.002$ & $4.82\pm 0.02$ & $0.253\pm 0.015$ & $-0.584\pm 0.003$ & $-4.42\pm 0.05$\\
  & $(12.25, 12.50]$ & $0.334\pm 0.012$ & $-0.587\pm 0.002$ & $4.89\pm 0.02$ & $0.333\pm 0.016$ & $-0.606\pm 0.004$ & $-4.70\pm 0.06$\\
  & $(12.50, 12.75]$ & $0.468\pm 0.016$ & $-0.611\pm 0.003$ & $4.85\pm 0.03$ & $0.506\pm 0.024$ & $-0.627\pm 0.004$ & $-5.20\pm 0.08$\\
  & $(12.75, 13.00]$ & $0.686\pm 0.018$ & $-0.648\pm 0.005$ & $4.71\pm 0.05$ & $0.821\pm 0.036$ & $-0.673\pm 0.006$ & $-6.25\pm 0.12$\\
  & $(13.00, 13.25]$ & $0.922\pm 0.036$ & $-0.676\pm 0.006$ & $4.42\pm 0.09$ & $1.094\pm 0.043$ & $-0.701\pm 0.007$ & $-7.09\pm 0.14$\\
  & $(13.25, 13.50]$ & $1.147\pm 0.037$ & $-0.708\pm 0.008$ & $4.17\pm 0.08$ & $1.231\pm 0.048$ & $-0.712\pm 0.008$ & $-7.49\pm 0.15$\\
  & $(13.50, 13.75]$ & $1.272\pm 0.043$ & $-0.717\pm 0.010$ & $3.96\pm 0.11$ & $1.309\pm 0.060$ & $-0.718\pm 0.011$ & $-7.73\pm 0.19$\\
  & $(13.75, 14.00]$ & $1.572\pm 0.096$ & $-0.772\pm 0.015$ & $3.70\pm 0.21$ & $1.529\pm 0.075$ & $-0.767\pm 0.016$ & $-8.63\pm 0.27$\\
  & $(14.00, \infty)$ & $1.588\pm 0.091$ & $-0.750\pm 0.019$ & $3.48\pm 0.27$ & $1.412\pm 0.068$ & $-0.749\pm 0.014$ & $-8.21\pm 0.23$\\
  \noalign{\smallskip}\hline
\end{tabular}
\end{center}
\tablefoot{
The environmental dependence of the FP and the FP$_*$ of central galaxies in $g$, $r$, $i$ and $z$ bands. Column 2 is the logarithmic halo mass, Cols.~3 to 5 are FP$_*$ coefficients, and Cols.~6 to 8 are FP coefficients.
}
\end{table*}

\begin{table*}[p]
\begin{center}
\caption[]{Environmental dependence of the FP and the FP$_*$ of satellite galaxies \label{tab:envsat}}
 \begin{tabular}{cccccccc}
  \hline\noalign{\smallskip}
Band & $log_{10}M_{halo}$ & \multicolumn{3}{c}{FP$_*$ Coefficients} & \multicolumn{3}{c}{FP Coefficients}\\
 & & $a$ & $b$ & $c$ & $a$ & $b$ & $c$\\
  \hline\noalign{\smallskip}
$g$ & $(12.00, 12.25]$ & $1.055\pm 0.137$ & $-0.650\pm 0.025$ & $3.69\pm 0.19$ & $0.965\pm 0.132$ & $-0.697\pm 0.027$ & $-7.30\pm 0.46$\\
  & $(12.25, 12.50]$ & $1.286\pm 0.072$ & $-0.697\pm 0.020$ & $3.63\pm 0.15$ & $1.185\pm 0.067$ & $-0.721\pm 0.019$ & $-7.96\pm 0.26$\\
  & $(12.50, 12.75]$ & $1.178\pm 0.047$ & $-0.672\pm 0.016$ & $3.65\pm 0.14$ & $1.067\pm 0.048$ & $-0.717\pm 0.014$ & $-7.67\pm 0.18$\\
  & $(12.75, 13.00]$ & $1.242\pm 0.033$ & $-0.670\pm 0.011$ & $3.51\pm 0.09$ & $1.160\pm 0.035$ & $-0.717\pm 0.009$ & $-7.85\pm 0.13$\\
  & $(13.00, 13.25]$ & $1.319\pm 0.033$ & $-0.693\pm 0.008$ & $3.54\pm 0.08$ & $1.204\pm 0.027$ & $-0.724\pm 0.009$ & $-8.01\pm 0.10$\\
  & $(13.25, 13.50]$ & $1.403\pm 0.028$ & $-0.714\pm 0.008$ & $3.54\pm 0.09$ & $1.278\pm 0.026$ & $-0.752\pm 0.008$ & $-8.40\pm 0.10$\\
  & $(13.50, 13.75]$ & $1.426\pm 0.019$ & $-0.730\pm 0.008$ & $3.62\pm 0.07$ & $1.279\pm 0.023$ & $-0.767\pm 0.006$ & $-8.53\pm 0.08$\\
  & $(13.75, 14.00]$ & $1.469\pm 0.020$ & $-0.744\pm 0.009$ & $3.64\pm 0.07$ & $1.297\pm 0.016$ & $-0.776\pm 0.007$ & $-8.65\pm 0.07$\\
  & $(14.00, \infty)$ & $1.469\pm 0.013$ & $-0.778\pm 0.006$ & $3.93\pm 0.05$ & $1.291\pm 0.011$ & $-0.795\pm 0.005$ & $-8.79\pm 0.05$\\
  \noalign{\smallskip}\hline
$r$ & $(12.00, 12.25]$ & $1.125\pm 0.143$ & $-0.683\pm 0.029$ & $3.84\pm 0.18$ & $1.040\pm 0.136$ & $-0.707\pm 0.029$ & $-7.33\pm 0.48$\\
  & $(12.25, 12.50]$ & $1.358\pm 0.073$ & $-0.738\pm 0.022$ & $3.84\pm 0.15$ & $1.261\pm 0.067$ & $-0.745\pm 0.020$ & $-8.08\pm 0.26$\\
  & $(12.50, 12.75]$ & $1.232\pm 0.047$ & $-0.713\pm 0.013$ & $3.90\pm 0.11$ & $1.132\pm 0.046$ & $-0.732\pm 0.014$ & $-7.69\pm 0.18$\\
  & $(12.75, 13.00]$ & $1.281\pm 0.033$ & $-0.704\pm 0.011$ & $3.73\pm 0.09$ & $1.208\pm 0.035$ & $-0.729\pm 0.010$ & $-7.82\pm 0.13$\\
  & $(13.00, 13.25]$ & $1.345\pm 0.033$ & $-0.727\pm 0.009$ & $3.79\pm 0.09$ & $1.252\pm 0.027$ & $-0.739\pm 0.009$ & $-7.99\pm 0.10$\\
  & $(13.25, 13.50]$ & $1.421\pm 0.028$ & $-0.744\pm 0.008$ & $3.77\pm 0.09$ & $1.324\pm 0.025$ & $-0.760\pm 0.008$ & $-8.32\pm 0.10$\\
  & $(13.50, 13.75]$ & $1.436\pm 0.019$ & $-0.767\pm 0.008$ & $3.93\pm 0.07$ & $1.326\pm 0.022$ & $-0.778\pm 0.007$ & $-8.47\pm 0.08$\\
  & $(13.75, 14.00]$ & $1.481\pm 0.019$ & $-0.780\pm 0.009$ & $3.94\pm 0.07$ & $1.350\pm 0.016$ & $-0.790\pm 0.007$ & $-8.62\pm 0.07$\\
  & $(14.00, \infty)$ & $1.476\pm 0.013$ & $-0.812\pm 0.005$ & $4.22\pm 0.05$ & $1.343\pm 0.011$ & $-0.808\pm 0.005$ & $-8.75\pm 0.05$\\
  \noalign{\smallskip}\hline
$i$ & $(12.00, 12.25]$ & $1.188\pm 0.148$ & $-0.707\pm 0.031$ & $3.92\pm 0.17$ & $1.103\pm 0.138$ & $-0.720\pm 0.031$ & $-7.46\pm 0.49$\\
  & $(12.25, 12.50]$ & $1.389\pm 0.077$ & $-0.753\pm 0.023$ & $3.91\pm 0.15$ & $1.303\pm 0.065$ & $-0.764\pm 0.020$ & $-8.20\pm 0.26$\\
  & $(12.50, 12.75]$ & $1.253\pm 0.048$ & $-0.730\pm 0.014$ & $4.01\pm 0.12$ & $1.169\pm 0.046$ & $-0.745\pm 0.014$ & $-7.77\pm 0.18$\\
  & $(12.75, 13.00]$ & $1.298\pm 0.034$ & $-0.720\pm 0.011$ & $3.84\pm 0.09$ & $1.235\pm 0.034$ & $-0.740\pm 0.010$ & $-7.86\pm 0.13$\\
  & $(13.00, 13.25]$ & $1.364\pm 0.033$ & $-0.744\pm 0.009$ & $3.90\pm 0.08$ & $1.281\pm 0.026$ & $-0.756\pm 0.010$ & $-8.08\pm 0.10$\\
  & $(13.25, 13.50]$ & $1.440\pm 0.028$ & $-0.761\pm 0.008$ & $3.88\pm 0.09$ & $1.352\pm 0.025$ & $-0.775\pm 0.008$ & $-8.38\pm 0.10$\\
  & $(13.50, 13.75]$ & $1.450\pm 0.019$ & $-0.779\pm 0.008$ & $4.01\pm 0.08$ & $1.357\pm 0.022$ & $-0.791\pm 0.007$ & $-8.52\pm 0.08$\\
  & $(13.75, 14.00]$ & $1.496\pm 0.019$ & $-0.797\pm 0.009$ & $4.05\pm 0.08$ & $1.380\pm 0.016$ & $-0.805\pm 0.007$ & $-8.68\pm 0.08$\\
  & $(14.00, \infty)$ & $1.495\pm 0.013$ & $-0.825\pm 0.005$ & $4.30\pm 0.05$ & $1.375\pm 0.011$ & $-0.821\pm 0.005$ & $-8.80\pm 0.05$\\
  \noalign{\smallskip}\hline  
$z$ & $(12.00, 12.25]$ & $1.129\pm 0.123$ & $-0.725\pm 0.028$ & $4.21\pm 0.16$ & $1.229\pm 0.142$ & $-0.754\pm 0.034$ & $-7.89\pm 0.52$\\
  & $(12.25, 12.50]$ & $1.315\pm 0.067$ & $-0.765\pm 0.021$ & $4.18\pm 0.14$ & $1.402\pm 0.071$ & $-0.790\pm 0.022$ & $-8.51\pm 0.28$\\
  & $(12.50, 12.75]$ & $1.215\pm 0.046$ & $-0.753\pm 0.014$ & $4.30\pm 0.11$ & $1.250\pm 0.050$ & $-0.762\pm 0.016$ & $-7.98\pm 0.20$\\
  & $(12.75, 13.00]$ & $1.265\pm 0.034$ & $-0.740\pm 0.014$ & $4.09\pm 0.11$ & $1.299\pm 0.035$ & $-0.761\pm 0.011$ & $-8.06\pm 0.14$\\
  & $(13.00, 13.25]$ & $1.323\pm 0.031$ & $-0.770\pm 0.009$ & $4.23\pm 0.08$ & $1.338\pm 0.027$ & $-0.778\pm 0.011$ & $-8.27\pm 0.12$\\
  & $(13.25, 13.50]$ & $1.414\pm 0.027$ & $-0.784\pm 0.009$ & $4.15\pm 0.09$ & $1.412\pm 0.026$ & $-0.790\pm 0.009$ & $-8.53\pm 0.10$\\
  & $(13.50, 13.75]$ & $1.431\pm 0.019$ & $-0.810\pm 0.008$ & $4.33\pm 0.07$ & $1.409\pm 0.022$ & $-0.809\pm 0.007$ & $-8.67\pm 0.08$\\
  & $(13.75, 14.00]$ & $1.490\pm 0.019$ & $-0.828\pm 0.009$ & $4.35\pm 0.07$ & $1.436\pm 0.017$ & $-0.827\pm 0.008$ & $-8.87\pm 0.08$\\
  & $(14.00, \infty)$ & $1.495\pm 0.013$ & $-0.853\pm 0.005$ & $4.55\pm 0.05$ & $1.430\pm 0.011$ & $-0.836\pm 0.005$ & $-8.93\pm 0.05$\\
  \noalign{\smallskip}\hline
\end{tabular}
\end{center}
\tablefoot{
The environmental dependence of the FP and the FP$_*$ of satellite galaxies in $g$, $r$, $i$ and $z$ bands. Columns are similar to Table \ref{tab:envcen}.
}
\end{table*}

We find the following results. Firstly, in the same halo bin, coefficient $a$ of FP$_*$ is systematically larger than that of the FP, except for in $z$ band. This means that the tilt of the FP$_*$ is indeed mitigated relative to that of the FP. Secondly, coefficient $b$ of the FP$_*$ is similar to that of the FP. Finally, the environmental dependence of the FP$_*$ is similar to that of the FP. Although FP$_*$ coefficients of central galaxies are correlated to the halo mass, those of satellite galaxies are similar in different halo mass bins. These results suggest that the FP tilt is not primarily driven by the stellar population. Moreover, the discrepancy between central and satellite galaxies is not due to the difference of the stellar population in them.

\begin{figure*}
\begin{center}
\includegraphics[scale=0.55]{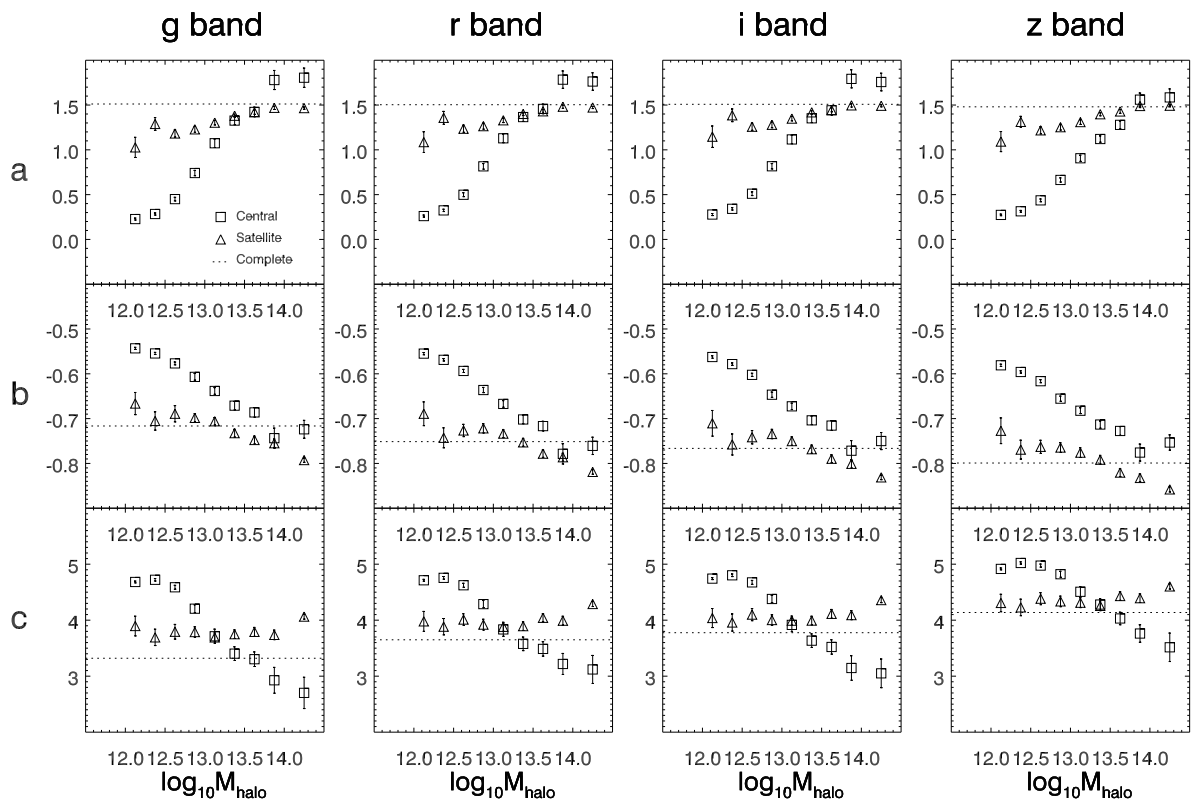}
\end{center}

\caption{As in Fig.~\ref{fig:fp_bin_star}, but galaxies are redder than the dividing line $g-r=0.68-0.030(M_r +21)$.}
\label{fig:fp_bin_star_red}
\end{figure*}

To test whether our results are due to the pollution of late-type galaxies in the sample, we select ETGs that are redder than the dividing line $g-r=0.68-0.030(M_r +21)$, and calculate the environmental dependence of the FP$_*$ of these galaxies. Results are shown in Fig.~\ref{fig:fp_bin_star_red}. We find that the pollution of blue galaxies does not change our conclusions. 

\begin{figure*}
\begin{center}
\includegraphics[scale=0.55]{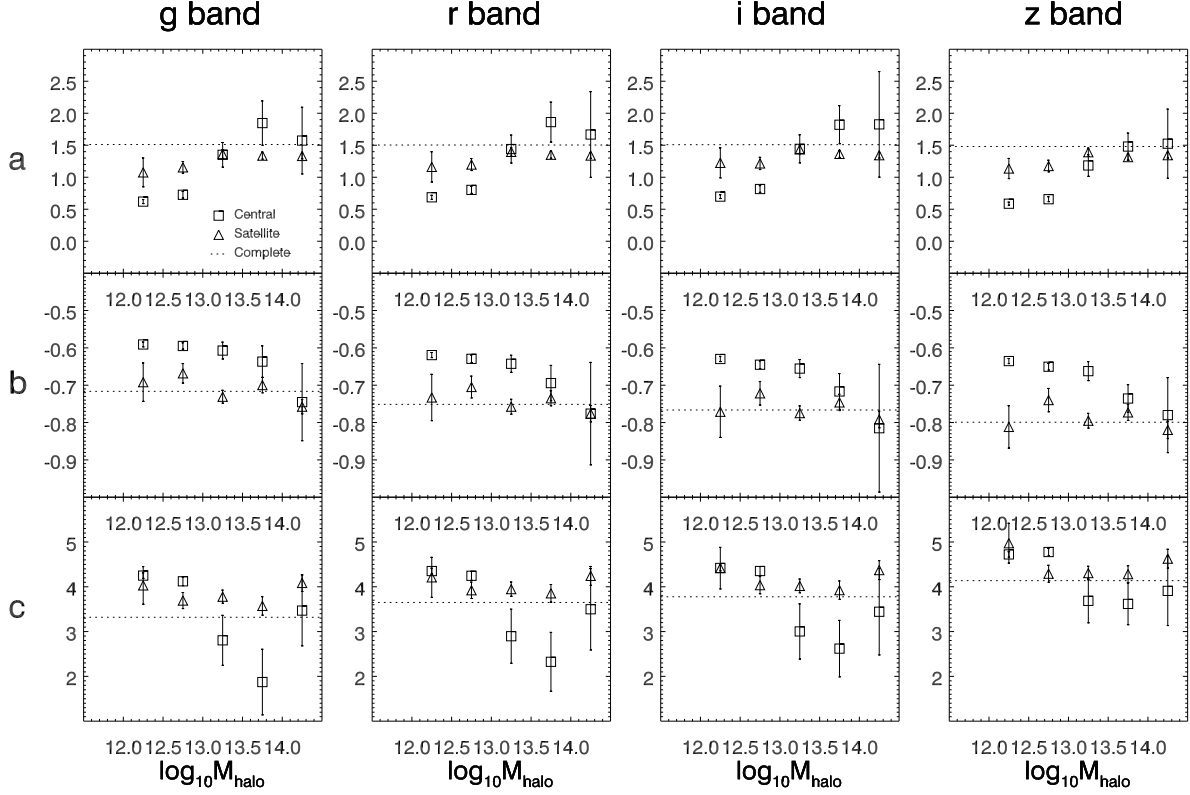}
\end{center}
\caption{As in Fig.~\ref{fig:fp_bin_star}, but galaxies in each halo mass bin are located in the same volume of the parameter space.}
\label{fig:fp_bin_star_volume}
\end{figure*}

The FP was considered as a universal scaling relation, in the sense that ETGs with different properties obey a similar FP. However, there is evidence that the scaling relation is curved when expanding the range of galaxy properties \citep{2006ApJ...638..725Z, 2010MNRAS.406.1220W, 2011ApJ...726..108T}. Therefore, when comparing FP coefficients of two samples, we should ensure that galaxies in these samples are located in the same volume of the parameter space. To do this, we fit the distribution of $\log R_0$, $\log \sigma_0$, and $\log M_*$ in each halo mass bin to the double Gaussian. Assuming the fitted peaks and standard deviations of $\log R_0$ distribution are $\mu_1$, $\sigma_1$, $\mu_2$, and $\sigma_2$, with $\mu_1<\mu_2$, we only keep the galaxies with $\log R_0$ in the range [$\mu_1$-$\sigma_1$, $\mu_2$+$\sigma_2$]. Similarly, galaxies are screened so that central and satellite galaxies are in the same range of $\log \sigma_0$ and $\log M_*$. Afterwards, the remaining galaxies in the same halo mass bin are located in the same volume of ($R_0$, $\sigma_0$ and $M_*$) space. We plot the environmental dependence of the FP$_*$ of the resulting samples in Fig.~\ref{fig:fp_bin_star_volume}. Here we set the size of the halo mass bin as 0.5, so that galaxy numbers in all halo mass bins are statistically large enough. We find that the environmental dependence of the FP$_*$ still exists. Therefore, the difference between central and satellite galaxies is not primarily due to the curvature of the scaling relation.

\section{Summary and discussion}
\subsection{Summary}
Using a sample of 70793 ETGs from SDSS DR7 and the SDSS group catalog, we investigate the environmental dependence of the FP$_*$. We find that coefficient $a$ of the FP$_*$ is closer to 2 than that of the FP, but coefficient $b$ of the FP$_*$ is similar to that of the FP. The environmental dependence of the FP$_*$ is similar to that of the FP. The FP$_*$ coefficients of central galaxies are correlated to the halo mass, but those of satellite galaxies are similar in all halo mass bins. These results suggest that the FP tilt cannot be explained by the stellar population, and that the difference between central and satellite galaxies is not due to the stellar population.

\subsection{Explanations of the results}

The similarity between the environmental dependence of the FP and the FP$_*$ implies that the discrepancy between central and satellite galaxies is not due to variations in the stellar mass-to-light ratio $M_*/L$. We discuss several explanations for the results as follows.

\subsubsection{Environmental difference between central and satellite galaxies}
FP$_*$ coefficents of central and satellite galaxies reveal different dependencies on the halo mass. One explanation is that central and satellite galaxies are affected by different environmental processes, which make them obey different FPs. There is a significant difference between the properties of central and satellite galaxies \citep[e.g.][]{2006MNRAS.366....2W, 2008MNRAS.387...79V, 2010ApJ...721..193P, 2012ApJ...757....4P, 2013MNRAS.428.3121M, 2013MNRAS.428.3306W, 2013MNRAS.432..336W}. In the standard $\Lambda$CDM cosmology, the central galaxy resides in the middle of the dark matter halo, where the density is higher and the gravitational potential well is deep, while satellite galaxies are considered as having been previously accreted into the halo and then into the orbit around the central galaxy. Several environmental processes only operate on satellite galaxies, such as ram pressure, tidal stripping, and strangulation. Therefore, the environmental difference between central and satellite galaxies drives them to evolve in different ways. Essentially, this enables the FPs of central and satellite galaxies to be made differently. Furthermore, the concentration of the dark matter halo is smaller in more massive halos \citep[e.g.,][]{1997ApJ...490..493N, 2001MNRAS.321..559B, 2001ApJ...554..114E, 2009ApJ...707..354Z}. As a result, the environmental difference between central and satellite galaxies is more significant in less massive halos. This is consistent with our results: FP$_*$ coefficients of central and satellite galaxies are similar in large halos, but are significantly different in small halos.

An alternative explanation is that the environment does not change the FP, but only affects how galaxies populate the FP. In other words, the FP is driven by a basic physical process, while galaxies in different environments are located on a different portion of the FP. It is difficult to conclude whether this explanation is more relevant, because the connection between the environment and the location on the FP may only be the consequence of the environmental process, mentioned above.

\subsubsection{Rotational component}

Traditionally, different scaling relations have been available to describe the kinematic and photometric properties of disk galaxies (such as the Tully-Fisher relation) and ETGs (such as the Faber-Jackson relation and the FP). In these relations, the kinematics of disk galaxies is described by the rotational velocity $v_c$, and that of ETGs is described by the velocity dispertion $\sigma$. However, several studies conclude that the scatter of scaling relations (such as the Tully-Fisher relation and the FP) is smaller by adopting $V\equiv\sqrt{v_c^2/2+\sigma^2}$ as the kinematic parameter \citep{1997AJ....114.1365B, 2006ApJ...653.1049W, 2007ApJ...660L..35K, 2008ApJ...682...68Z}. \citet{2008ApJ...682...68Z} studied galaxies from disks to spheroids and from dwarf spheroidals to brightest cluster galaxies, and found that all galaxies obey the following equation, called the fundamental manifold (FM):
\begin{equation}
\log r_e-\log V^2+\log I_e+\log\Upsilon_e+0.8=0,
\label{equa:curve}
\end{equation}
where $r_e$ is the half-light radius, $I_e$ and $\Upsilon_e$ are the surface brightness and the mass-to-light ratio within $r_e$ respectively. $\Upsilon_e$ includes the contribution of the stellar-to-baryonic mass ratio and the concentration of stars in the halo. Therefore, the scaling relation between the size, kinematics, and luminosity is explained completely by the virial theorem and the behavior of $\Upsilon_e$.

When deriving Equations \ref{equa:virial} and \ref{equa:star}, we assume $V$ is proportional to $\sigma$. This is challenged by the fact that $v_c/\sigma$ depends on galaxy properties, such as the luminosity, the color and the isophotal shape of the surface brightness \citep[e.g.,][]{1983ApJ...266...41D, 1996ApJ...464L.119K}. There is evidence that satellite galaxies have systematically higher rotational velocity than central galaxies \citep{2015MNRAS.454..322D}. This may also be responsible for the environmental dependence of the color, because bluer galaxies are likely to have a larger rotational component. However, the SDSS data do not provide the data for rotational velocity. Assuming $V\propto \sigma^\gamma$, the existence of the rotational component would make $\gamma<1$, and lead to FP coefficient $a<2$. If $\gamma$ of satellite galaxies is systematically larger than that of central galaxies, coefficent $a$ would be larger for satellite galaxies. Therefore, the environmental dependence of the FP may be due to the missing rotation component in the formulation.

\subsubsection{Differences in the parameter volume}

In Section 3, we keep central and satellite galaxies located in the same parameter volume, and find that this correction would not change our conclusions. However, a complete correction should consider the fact that central and satellite galaxies are located in different parameter volumes, which is still an environmental result. The study on central and satellite galaxies, with similar ditributions, would help to solve this problem, but a much larger sample would be required.

\subsubsection{Variations in the IMF}

When calculating the stellar mass, we assume that all ETGs have a Kroupa IMF. \citet{2001ApJ...550..212B} find that, if the IMF is universal, there is a linear relation between the stellar mass-to-light ratio in the band $\lambda$ and the color
\begin{equation}
\log(M_*/L) = a_\lambda+b_\lambda \times color.
\end{equation}
The slope $b_\lambda$ is insensitive to the assumed IMF, but the normalization $a_\lambda$ is strongly affected by the IMF. Therefore, changing the Kroupa IMF to another universal IMF produces a constant offset to the estimation of $\log M_*$, and thus only produes a constant offset to the coefficient $c$.

\begin{figure*}
\begin{center}
\includegraphics[scale=0.55]{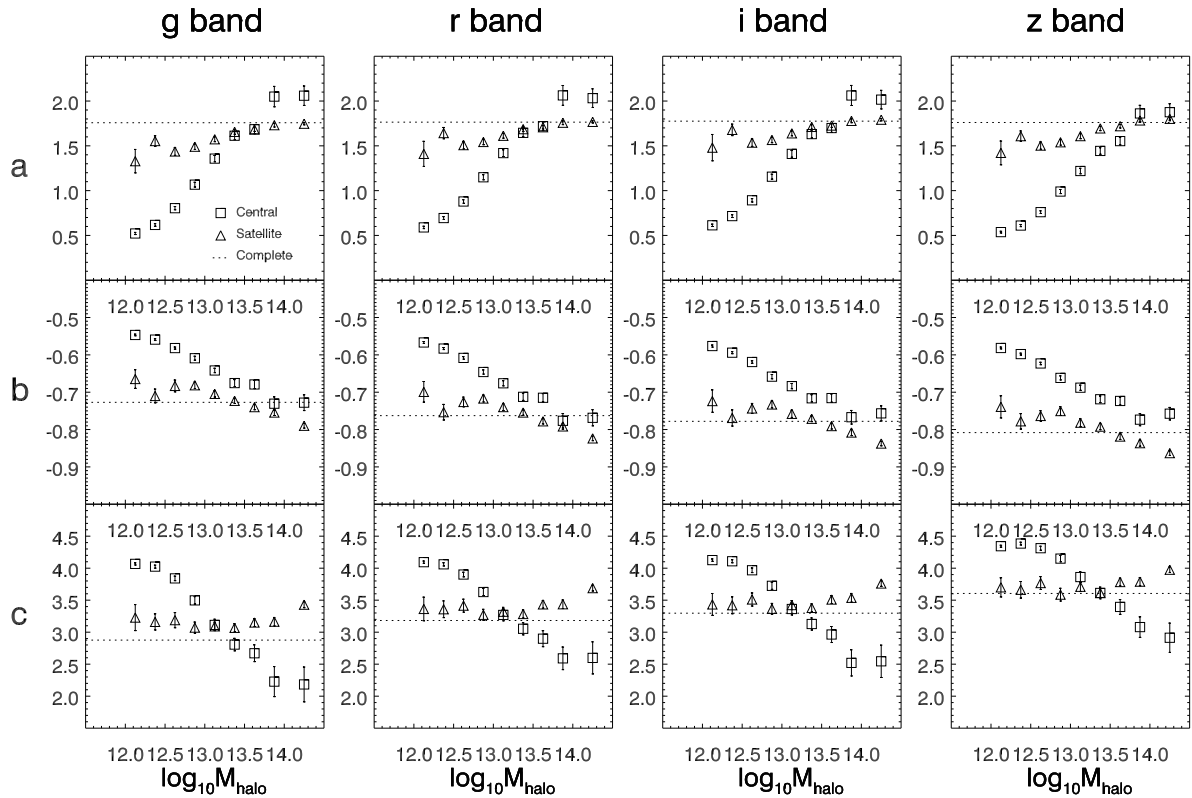}
\end{center}
\caption{As in Fig.~\ref{fig:fp_bin_star}, but the stellar mass $M_*$ is corrected to take account of the IMF variations. The correction is based on the relation between the stellar mass and the velocity dispersion, which is parameterized by \citet{2013MNRAS.432.1862C}.}
\label{fig:fp_bin_star_corr}
\end{figure*}

However, there is evidence that the IMF is not universal but correlated to the properties of ETGs \citep[e.g.,][]{2003MNRAS.339L..12C, 2010ApJ...709.1195T, 2010Natur.468..940V, 2012Natur.484..485C, 2012ApJ...753L..32S, 2013MNRAS.432.1862C, 2015MNRAS.446..493P}. This is described as the IMF mismatch parameter $\alpha \equiv M_{stars}/M_{pop}$, where $M_{stars}$ is the stellar mass that has been calculated using the lensing and dynamical model, and $M_{pop}$ is the stellar mass that has been calculated using the stellar populations synthesis models with the universal IMF. According to \citet{2013MNRAS.432.1862C}, $\alpha$ is correlated to the velocity dispersion as
\begin{equation}
\log\alpha = a'+b'\log\sigma_0.
\end{equation}
To test the effect of the IMF on our conclusions, we make a correction to the stellar mass as
\begin{equation}
\log M_{*,corrected}=\log M_*+\log\alpha,
\end{equation}
and use the corrected stellar mass to calculate the environmental dependence of the FP$_*$, as illustrated in Fig.~\ref{fig:fp_bin_star_corr}. We note that, after the correction on the IMF, coefficent $a$ is systematically larger, coefficient $b$ is not affected, and coefficent $c$ is systematically smaller. Therefore, the correction on the IMF decreases the tilt of the FP$_*$, and does not change the environmental dependence.

\subsubsection{Non-homology}

Several works indicate that the structural and dynamical non-homology plays an important role in the FP tilt \citep{2004ApJ...600L..39T, 2006MNRAS.370..681N, 2010MNRAS.408.1335L, 2013MNRAS.435...45D}. However, the SAURON project studied the dynamical properties of ETGs, and concluded that $M_{tot}/L$ can explaine most of the FP tilt and that the contribution of non-homology is little \citep{2006MNRAS.366.1126C,2013MNRAS.432.1709C}. \citet{2008ApJ...684..248B} used the gravitational lensing to estimate the total mass, and also concluded that the non-homology is not important. Therefore, the contribution of the non-homology is still unclear.

\subsubsection{Dark matter fraction}
 
\citet{2006MNRAS.366.1126C} indicate the dark matter fraction of slow-rotating galaxies is higher than that of fast-rotating galaxies. \citet{2009MNRAS.396.1132T} found that the dark matter fraction depends on the luminosity of galaxies. Moreover, recent observations and simulations indicated the baryonic fraction is correlated to the halo mass \citep{2009ApJ...703..982G,2011ApJ...743...13L,2012ApJ...745L...3L, 2012NJPh...14d5004S,2013JCAP...04..022B,2014MNRAS.442.2641V}. Variations in the dark matter fraction may be responsible for the environmental dependence.

\subsection{The FP as a distance indicator}
The FP provides an estimation of the size of ETGs if the velocity dispersion and surface brightness are known, and thus can be used as a distance indicator. As the extension of the FP, the fundamental manifold (Equation \ref{equa:curve}) is also used as a fiducial for other distance indicators \citep{2012ApJ...748...15Z}. However, the environmental dependence of the FP and the FP$_*$ indicates that there is a systematic bias between galaxies in different environments. The effect of the environment should be considered when scaling relations are used to estimate the distance of galaxies. If the environmental dependence of the FP is due to the incomplete correction for differences in parameter volumes, in other words, the results in this work are only the consequence of the curvature of the scaling relation, the distance estimation would not be affected.

\begin{acknowledgements}
We are grateful to Xiaohu Yang for providing the SDSS group catalog. WY would like to acknowledge the support of the Fundamental Research Funds for the Central Universities, 973 Program (N0. 2015CB857004), the Science Fund for Creative Research Groups of the National Natural Science Foundation of China (Grant NSFC 11421303), and Nature Science Foundation of Anhui Province.
\end{acknowledgements}

\bibliographystyle{aa}
\bibliography{stellarfp}

\end{document}